\documentclass[conference]{IEEEtran}

\usepackage[T1]{fontenc}
\usepackage[utf8]{inputenc}
\usepackage{cite}
\usepackage{amsmath,amssymb,amsfonts}
\usepackage{algorithmic}
\usepackage{graphicx}
\usepackage{textcomp}
\usepackage{bmpsize}
\usepackage{xcolor}
\usepackage{lipsum}
\usepackage[colorlinks=true,urlcolor=black]{hyperref}
\usepackage{bm}
\usepackage{url}
\usepackage{tikz}
\usepackage{csquotes}
\usepackage{todonotes}
\usepackage{wasysym}
\usepackage{listings}
\usepackage{enumitem}
\usepackage[capitalise,noabbrev]{cleveref}
\crefname{paragraph}{Paragraph}{Paragraphs}
\usepackage{xspace}

\def\BibTeX{{\rm B\kern-.05em{\sc i\kern-.025em b}\kern-.08em
		T\kern-.1667em\lower.7ex\hbox{E}\kern-.125emX}}

\usepackage{acro}
\DeclareAcronym{TAN}{
	short = TAN,
	long = transaction authentication number,
}
\DeclareAcronym{iTAN}{
	short = iTAN,
	long = indexed TAN,
}
\DeclareAcronym{OTP}{
	short = OTP,
	long = one-time password,
}
\DeclareAcronym{TOTP}{
	short = TOTP,
	long = time-based one-time password
}
\DeclareAcronym{HOTP}{
	short = HOTP,
	long = HMAC-based one-time password
}
\DeclareAcronym{CAP}{
	short = CAP,
	long = chip authentication program
}
\DeclareAcronym{WebAuthn}{
	short = WebAuthn,
	long = Web Authentication
}
\DeclareAcronym{W3C}{
	short = W3C,
	long = World Wide Web Consortium
}
\DeclareAcronym{CTAP}{
	short = CTAP,
	long = Client to Authenticator Protocol
}
\DeclareAcronym{U2F}{
	short = U2F,
	long = Universal Second Factor
}
\DeclareAcronym{txAuthSimple}{
	short = txAuthSimple,
	long = Simple Transaction Authorization Extension
}
\DeclareAcronym{ATS}{
	short = ATS,
	long = automatic transfer system
}
\DeclareAcronym{mTAN}{
	short = mTAN,
	long = mobile TAN
}
\DeclareAcronym{AAL}{
	short = AAL,
	long = Authenticator Assurance Level
}
\DeclareAcronym{PSD2}{
	short = PSD2,
	long = revised Payment Services Directive
}
\DeclareAcronym{2DA-Blueprint}{
	short = 2DA,
	long = Two Display Authentication
}
\DeclareAcronym{2DA-Scheme}{
	short = FIDO2D,
	long = FIDO2 With Two Displays
}
\DeclareAcronym{tamarin}{
	short = Tamarin,
	long = Tamarin Prover
}
\DeclareAcronym{TLS}{
	short = TLS,
	long = Transport Layer Security
}
\DeclareAcronym{TEE}{
	short = TEE,
	long = Trusted Execution Environment
}
\DeclareAcronym{UC}{
	short = UC,
	long = Universal Composability
}
\DeclareAcronym{FIDO2}{
	short = FIDO2,
	long = FIDO2
}
\DeclareAcronym{PIN}{
	short = PIN,
	long = Personal Identification Number
}
\acuse{FIDO2}
\DeclareAcronym{FIDO} {
	short = FIDO,
	long = Fast IDentity Online
}
\DeclareAcronym{RBA} {
	short = RBA,
	long = risk-based authentication
}

\newcommand{\trl}[1]{\ensuremath{\mathsf{#1}}}

\lstdefinelanguage{tamarin}{
	morekeywords={rule,lemma,All,Ex,theory,begin,builtins,end,exists-trace,restriction,ifdef,endif},
	alsoletter={-},
	morecomment=[l]{//}
}

\expandafter\def\expandafter\UrlBreaks\expandafter{\UrlBreaks
	\do\a\do\b\do\c\do\d\do\e\do\f\do\g\do\h\do\i\do\j%
	\do\k\do\l\do\m\do\n\do\o\do\p\do\q\do\r\do\s\do\t%
	\do\u\do\v\do\w\do\x\do\y\do\z\do\A\do\B\do\C\do\D%
	\do\E\do\F\do\G\do\H\do\I\do\J\do\K\do\L\do\M\do\N%
	\do\O\do\P\do\Q\do\R\do\S\do\T\do\U\do\V\do\W\do\X%
	\do\Y\do\Z}

\begin{document}

\title{FIDO2 With Two Displays---Or How to Protect Security-Critical Web Transactions Against Malware Attacks}

\author{\IEEEauthorblockN{Timon Hackenjos\IEEEauthorrefmark{1}, Benedikt Wagner\IEEEauthorrefmark{2}, Julian Herr\IEEEauthorrefmark{1}, Jochen Rill\IEEEauthorrefmark{3}, Marek Wehmer\IEEEauthorrefmark{1}, Niklas Goerke\IEEEauthorrefmark{1}, Ingmar Baumgart\IEEEauthorrefmark{1}}
	\IEEEauthorblockA{\IEEEauthorrefmark{1}FZI Research Center for Information Technology}
	\IEEEauthorblockA{\IEEEauthorrefmark{2}CISPA Helmholtz Center for Information Security}
	\IEEEauthorblockA{\IEEEauthorrefmark{3}Alter Solutions Deutschland}
}
	
\maketitle

\begin{abstract}
	With the rise of attacks on online accounts in the past years, more and more services offer two-factor authentication for their users.
	Having factors out of two of the three categories \emph{something you know}, \emph{something you have} and \emph{something you are} should ensure that an attacker cannot compromise two of them at once.
	Thus, an adversary should not be able to maliciously interact with one's account.
	However, this is only true if one considers a weak adversary.
	In particular, since most current solutions only authenticate a \emph{session} and not individual \emph{transactions}, they are noneffective if one's device is infected with malware.
	For online banking, the banking industry has long since identified the need for authenticating transactions.
	However, specifications of such authentication schemes are not public and implementation details vary wildly from bank to bank with most still being unable to protect against malware.
	In this work, we present a generic approach to tackle the problem of malicious account takeovers, even in the presence of malware.
	To this end, we define a new paradigm to improve two-factor authentication that involves the concepts of \emph{one-out-of-two security} and \emph{transaction authentication}.
	Web authentication schemes following this paradigm can protect security-critical transactions against manipulation, even if one of the factors is completely compromised.
	Analyzing existing authentication schemes, we find that they do not realize one-out-of-two security.
	We give a blueprint of how to design secure web authentication schemes in general.
	Based on this blueprint we propose \ac{2DA-Scheme}, a new web authentication scheme based on the \acs{FIDO2} standard and prove its security using \acs{tamarin}.
	We hope that our work inspires a new wave of more secure web authentication schemes, which protect security-critical transactions even against attacks with malware.
\end{abstract}
\acresetall


\section{Introduction}
In recent years, the World Wide Web and the multitude of different services offered within changed almost everyone's life.
Whether we want to send an email, do online banking, buy a product, or update our social media profile, we use the Web.
In general, each of these activities requires a separate account with which we authenticate ourselves.
According to a study by Google in partnership with The Harris Poll in 2019, the average American has 27 different online accounts~\cite{harrispoll2019}.

Passwords always were and still are the main means of authentication on the Internet.
However, since one cannot possibly remember that many different passwords (and the adoption of password managers is still very low) re-using passwords on a multitude of different sites is a common practice~\cite{das2014tangled}.
This means that one stolen password from an account on an unimportant website might also give an adversary access to more important ones, like an online banking account.

As more and more of our private and professional life started to happen on the Internet, compromising accounts through simple and cheap attacks like credential stuffing and phishing became increasingly more attractive and lucrative for attackers.

To mitigate these risks, security experts started to advertise the use of two-factor authentication instead of only a single password~\cite{ion2015no}.
A mandatory second authenticating factor (like a \ac{OTP}) severely restricts the utility of stolen passwords for an attacker.
Users and account providers alike took a very long time to adopt this recommendation, but nowadays many web applications offer at least one form of two-factor authentication.
Most implementations require the user to present a second factor together with a password during login.
If successful, the user then receives an authenticated session token (stored within a cookie), which she can use to interact with the site and her account without having to authenticate herself again for a while.

When using two-factor authentication, it is recommended to use two of the three following factors: something you know, something you have and something you are.
Behind this categorization lies the assumption that it is improbable for the same adversary to compromise factors from different categories (e.g., an adversary that gets a password from a password leak cannot also steal a copy of the user's fingerprint).
However, this is only true in a very specific adversarial model.
An attacker who has compromised a victim's computer does not need to steal a copy of the fingerprint; he can simply manipulate all interactions with a website by using the authenticated session token stored in the browser after the victim performed a legitimate login.

Indeed, two-factor authentication as it is used today does not help against a number of attack techniques used by real-life adversaries~\cite{rsamitb2010, bestpracticesamnesty2018, zdnet2fabypass2019}.
Most schemes are susceptible to malware attacks and some popular forms, like \ac{OTP}, are also vulnerable to real-time phishing, in which an adversary relays authentication details from a fake website to a legitimate one~\cite{jacomme2021extensive, konoth2020securepay}.

Bruce Schneier adequately summarized the applicability of two-factor authentication in as early as 2005: \enquote{Two-factor authentication isn't our savior.
	It won't defend against phishing.
	It's not going to prevent identity theft.
	It's not going to secure online accounts from fraudulent transactions.
	It solves the security problems we had 10 years ago, not the security problems we have today}~\cite{schneier2005two}.

In recent years, online banking has been moving into the right direction security-wise by introducing \emph{transaction authentication}, which is the verification of transaction details (often on an additional device, but not necessarily).
Authenticating transactions individually mitigates the risk of session hijacking by stealing a cookie.
Furthermore, manipulation of transaction details can be detected if one is using an additional device to check them.

The \ac{CAP} protocol used in online banking thus provides a high level of security.
It relies on a dedicated card reader with a display.
The device offers a very small attack surface for infection by malware as it has limited functionality and is specifically built for this use case.

Carrying an additional device has been identified as unpleasant for many users~\cite{krol2015they, lyastani2020fido2}.
In consequence, many banks abandoned dedicated hardware tokens and transitioned to app-based authentication schemes such as photoTAN.
However, similar to regular computers, smartphones have large attack surfaces, are susceptible to, and have been attacked by malware~\cite{felt2011survey}.
By compromising the smartphone, attackers can bypass confirmation of transactions on the device~\cite{haupert2018app}.
Therefore, if an attacker gets access to the user's password, he can access the online banking system and execute arbitrary transactions (see \cref{sec:problems}).
Some banks even allow initiating transactions from the same device~\cite{haupert2018app}.
Thus, most current online banking schemes, which are required by law to offer strong security~\cite{PSD22017}, do not adequately protect against malware attacks, since the smartphone needs to be fully trusted.
Besides online banking, many other use cases such as administration panels and electronic health records handle security-critical transactions (see \cref{sec:usecases}).
However, most of them do not implement \emph{transaction authentication} and thus do not protect against malware attacks.

The recent \acs{FIDO2} standard provides a widely implemented browser API called \acs{WebAuthn} that simplifies integration of secure web authentication mechanisms relying on public-key cryptography and supporting authentication of individual transactions.
However, authentication with \acs{FIDO2} is primarily used for login today and thus suffers from susceptibility to malware attacks as described before.
The support for \emph{transaction authentication} has even been removed from the latest version of the \acs{WebAuthn} standard\cite{webauthn}.

We show how to design web authentication schemes using two devices (one of which could be a smartphone) which protect security-critical transactions, even if one device is fully compromised.
While this might seem like an impossible task, it can be done. In the following, we show how.

\section{Our Contribution}
In this work, we analyze the security of web authentication schemes in the presence of malware and real-time phishing attacks.
We find that even strong authentication schemes for online banking that rely on \emph{transaction authentication} as required by the \ac{PSD2} do not protect against these attacks.

As a remedy, we propose that web authentication schemes handling security-critical transactions should fulfill \emph{one-out-of-two security}, a security notion that neither requires a primary device nor an additional device trusted.
We show how this security notion introduced for electronic payment can be adapted to web authentication.
Furthermore, we provide a blueprint to design web authentication schemes that fulfill this notion and thus protect security-critical transactions even if one device is fully compromised.

Based on our blueprint, we design and implement \ac{2DA-Scheme}, a web authentication scheme based on the \ac{FIDO2} standard.
We identify shortcomings of \ac{FIDO2} for implementing malware-resistant web authentication and show how they can be treated.
Integration into the \ac{FIDO2} standard would pave the way for broad adoption of our approach.
We examine multiple use cases that benefit from the security guarantees of \ac{2DA-Scheme}.

Finally, we provide a formal model of \ac{2DA-Scheme} and prove that it fulfills \emph{one-out-of-two security} using \acs{tamarin}.

\acresetall
\acuse{FIDO2}
\section{Attacks on Web Authentication}
\label{sec:problems}
In this section, we analyze attacks on current web authentication schemes and introduce our attack model that serves as the basis for all of the following.

\subsection{Password Attacks}
\label{ssec:pw-attacks}
Password authentication is by far the most prominent authentication scheme in the web.
The main problems of password authentication are that users choose weak passwords and reuse them across multiple sites~\cite{thomas2017data}.
Brute-force and password spraying attacks exploit weak passwords while credential-stuffing attacks focus on reused passwords.
For example, the video conferencing solution Zoom was hit by a credential-stuffing attack and a large number of accounts were compromised~\cite{zoomcredstuffing2020}.
The success of these attacks is not surprising considering the amount of publicly available credentials~\cite{hunt2017password}.

\Ac{RBA} aims to strengthen password authentication by monitoring features such as the IP address and the user agent of the browser and triggering additional authentication during login if they differ from those recorded before~\cite{wiefling2021s}.
While this limits the impact of a stolen password, most features are easy to detect and spoof during a phishing or malware attack.
Criminal platforms evolved that sell access to user profiles containing credentials and features that allow bypassing \ac{RBA}~\cite{campobasso2020impersonation}.

\subsection{Real-Time Phishing}
\label{ssec:phishing}
In 2018, Amnesty International reported targeted phishing attacks on Google and Yahoo accounts that bypassed \ac{OTP}-based two-factor authentication~\cite{bestpracticesamnesty2018}.
By automating the process of using the stolen password and \ac{OTP}, the attackers were able to work around the short validity of the token.
Despite being more complex than classic phishing attacks that only harvest passwords for later use, real-time phishing attacks are not sophisticated.
Several tools are publicly available to automate this attack~\cite{evilginx2019, muraena2019}.
In addition, real-time phishing attacks can be used to identify features of the user necessary to bypass \ac{RBA}~\cite{campobasso2020impersonation}.

\subsection{Malware}
\begin{figure}[!t]
	\centering
	\includegraphics[width=\linewidth]{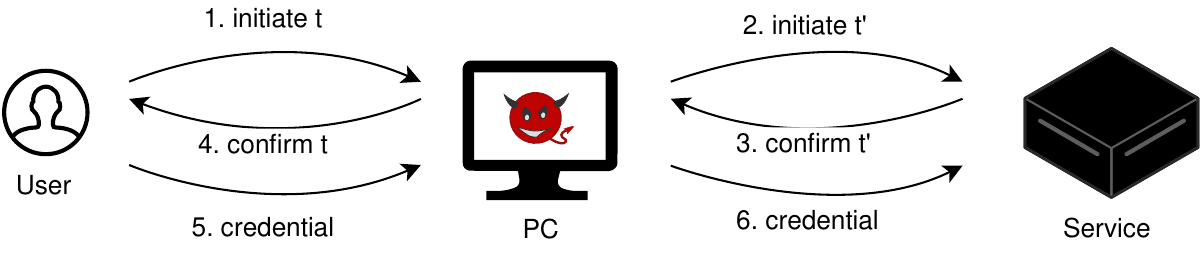}\hfill
	\caption{Transaction manipulation attack: malware on the device manipulates the transaction data t and issues a transaction t' instead.}
	\label{fig:attacks:malware}
\end{figure}
\begin{figure}[!t]
	\centering
	\includegraphics[width=\linewidth]{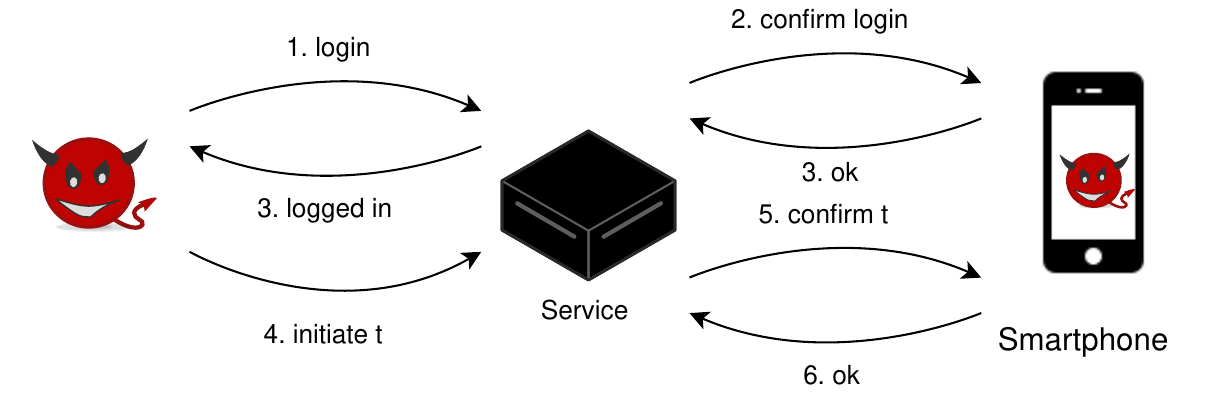}\hfill
	\caption{Transaction initiation attack: malware on a smartphone confirms transaction initiated by the attacker.}
	\label{fig:attacks:mobile-malware}
\end{figure}
As described in prior work, multiple ways exist to remotely infect devices with malware ranging from drive-by downloads and email attachments to social engineering attacks that convince a user to install malware herself~\cite{grier2012manufacturing, nelms2016towards}.
Google's Threat Analysis Group even detected the use of a zero-day exploit to install malware stealing cookies for popular websites such as Google, Microsoft and LinkedIn~\cite{tag2021zerodaycookies}.
Of course, smartphones are affected by malware too~\cite{felt2011survey}.
In the following, we discuss two types of attacks that can be carried out by malware to issue a malicious transaction.

\subsubsection{Transaction Manipulation}
In the following, we assume that an authentication scheme is used that does not display transaction details on an additional device.
Malware on the primary device can then carry out a transaction manipulation attack as depicted in \cref{fig:attacks:malware}.
Suppose that a user wants to make a security critical transaction on a website.
After logging in (potentially with two factors), she enters the desired transaction details t~(1), however, the compromised device initiates a manipulated transaction t'~(2).
If the service relies on session authentication only, the manipulated transaction t' is authenticated by the cookie sent by the browser and thus confirmed immediately.
However, the service might require confirmation by a second factor for transaction t' such as an \ac{OTP}~(3).
In this case, the malware prompts the user to confirm the original transaction t~(4).
As the user cannot detect the manipulation, she supplies the required \ac{OTP}~(5).
The attacker can use the \ac{OTP} to confirm the manipulated transaction t'~(6).
This attack applies to other authentication schemes such as \ac{FIDO2} as well.
Transaction manipulation has been used extensively in the wild by online banking malware such as ZeuS and SpyEye~\cite{rsamitb2010}.
Note that \ac{RBA} does not prevent this attack, as the transaction is initiated in a valid session of the user.

\subsubsection{Transaction Initiation}
Authentication schemes that incorporate an additional device for transaction verification are often susceptible to a transaction initiation attack as depicted in \cref{fig:attacks:mobile-malware}.
We assume that the attacker compromised the additional device and installed malware.
In a next step, the attacker needs to determine the user's password using one of the methods from \cref{ssec:pw-attacks,ssec:phishing}.
Potentially, the password is even stored on the additional device and thus accessible to the attacker.
With the password, the attacker logs in to the service from his own device using the victim's account~(1).
The service might require confirmation of the login attempt on the additional device, e.g., because the service implements \ac{RBA} and detected a change in the user profile.
Usually, confirmation of the login attempt relies on text messages, email or an app~\cite{wiefling2019really}.
However, since the attacker remotely controls the additional device, he can confirm the login request~(3)~\cite{haupert2018app}.
After logging in successfully, the attacker can initiate an arbitrary transaction t~(4).
The service might require confirmation of the transaction~(5), however, again the attacker controlling the additional device can confirm the request~(6).

\subsection{Our Attack Model}
We identify the following three main means of attack on web authentication schemes:
\begin{itemize}
	\item Password attacks
	\item Real-time Phishing
	\item Malware
\end{itemize}

Malware attacks are more powerful than real-time phishing and password attacks.
Password attacks give access to passwords only.
In addition, real-time phishing grants access to other credentials entered by the user such as an \ac{OTP}.
The strongest type of attack is compromising a device and infecting it with malware.
This grants full access to input and output interfaces of a device such as the keyboard and display, as well as credentials stored on the device.
Thus, malware on a device might eavesdrop on credentials entered by the user, as well as manipulate information displayed on the screen.

We assume malware to be executed with the highest user privileges available such as root or administrator and thus malware might manipulate the operating system.
This assumption is supported by prior work that documents the use of privilege escalation attacks in the wild~\cite{felt2011survey}.
Furthermore, a similar model for malware has been used in prior work~\cite{konoth2020securepay, haupert2018app}.

Because of the high prevalence of password and phishing attacks~\cite{thomas2017data}, we assume all attackers to be able to carry them out.
However, we differentiate attackers by their ability to infect involved devices with malware (see~\cref{sec:related-work}).
Furthermore, we consider that an attacker can carry out multiple attacks targeting the same user.
For example, an attacker that infects a smartphone with malware can also mount a password or phishing attack targeting the account of the user.
\section{Malware-Resistant Web Authentication}
\label{sec:paradigm}

As discussed in the last section, a secure web authentication scheme should protect against real-time phishing and malware attacks.
In the following, we argue that the concept of \emph{transaction authentication} is inevitable to protect against malware attacks and that \emph{two-factor authentication}, as it is currently defined, does not provide security benefits in this attack model.

\subsection{Transaction Authentication}
Usually, users \emph{sign in} into a web service before using it.
During this process, the web service authenticates the user and establishes a session.
\emph{Session authentication} means that that once the session is created, the user is allowed to perform an arbitrary number of actions by sending requests to the service.
It can be described as authenticating the session and then assuming every action performed as part of the session has been authorized by the user.

On the other hand, using \emph{transaction authentication} transactions are initiated in an authenticated session, but confirmation of each individual transaction is required (e.g., verification of transaction details by the user on a separate device).
Nonetheless, \emph{transaction authentication} is usually used for security-critical transactions only.
For example in online banking, transferring money between a checking and a savings account of the same user could rely on \emph{session authentication}.
Security-critical transactions such as bank transfers to an external account should require \emph{transaction authentication}.
These transactions must be identified for each use case on its own.
We provide examples for such transactions in \cref{sec:usecases}.

Relying on \emph{session authentication} only is fundamentally flawed in the presence of malware.
Taking control of the session lets an attacker interact with the service on behalf of the user, independently of the used authentication procedure.
Thus, strong security guarantees can only be stated for transactions protected by \emph{transaction authentication}.
However, \emph{transaction authentication} is not sufficient to protect against malware attacks on its own~\cite{haupert2018app}.
Combined with other measures proposed below, individual transactions can be protected to make it impossible for an attacker to perform an unsolicited transaction without the user explicitly authenticating it.

\subsection{Flaws in Two-Factor Authentication}
The notion of two-factor authentication is too weak in a realistic attack model and does not imply any security guarantees in the presence of a malware or phishing attacker, even for individually authenticated transactions.
Intuitively, we expect a two-factor authentication scheme to be secure as long as one of the two factors is not compromised.
However, this is satisfied neither by the notion nor by currently deployed schemes in the presence of malware.
As described in \cref{sec:problems}, malware and real-time phishing attacks are not prevented thoroughly by current two-factor authentication schemes.

Two-factor authentication schemes used in online banking such as the \ac{CAP} protocol and photoTAN provide resistance against malware on the device running the browser.
However, the reason for this does not lie in the fact that they fulfill the two-factor authentication definition, but that they require the user to verify transaction details on a separate device.
Thus, we need a new way of defining two-factor authentication that fulfills the intuitive promise of the notion: ensuring the security of a scheme even if one of two factors is fully compromised.
This redefinition of two-factor authentication makes malware and real-time phishing attacks infeasible and thus solves the weaknesses of current web authentication schemes.

\subsection{One-out-of-two Security for the Web}
\label{one-out-of-two-security}
In their work on secure electronic payment, Achenbach et al.~\cite{achenbach2019your} formalize exactly this requirement.
The authors model an electronic payment protocol using two separate devices that is still secure (with regards to specific security properties) even if one of the devices is fully compromised.
They call this property \emph{one-out-of-two security} and show that it can be fulfilled by requiring the user to verify the transaction details on both devices.
We adapt this idea of one-out-of-two security to web authentication by letting the user verify details of individual transactions on two separate devices.

The authors rely on a so-called \emph{confirmation channel} to build a scheme that fulfills one-out-of-two security.
The transaction details are displayed to the user (e.g., on an additional device) and explicit confirmation is required before performing the transaction.
We reuse the idea of a confirmation channel in the design of our web authentication scheme.
However, whereas Achenbach et al. explicitly allow an attacker to break their protocol by guessing the \ac{PIN} correctly, we use a slightly stronger version of one-out-of-two security.
Our notion also prevents attacks that circumvent compromising a device by breaking an underlying authentication scheme such as passwords.
According to our definition, a protocol that uses two separate devices exhibits one-out-of-two security, if an attacker who has the capabilities defined in our attack model in \cref{sec:problems} is not able to issue a fraudulent transaction without compromising both devices.

Note that the goal of the security property is authenticity of transactions and not confidentiality of the transaction details.
To be able to confirm the authenticity of a transaction on two devices, the user must either see or enter the details on each device separately.
In either case, both devices will need to handle the transaction details in plain text.
Thus, compromising one device is enough to violate the confidentiality of transaction details.
As described in \cref{sec:usecases}, by limiting access to actions that return or change data on the service, the authenticity guarantee of transactions can be used to achieve \emph{integrity} and \emph{confidentiality} of data stored on the server.
\subsection{Blueprint for Secure Web Authentication}
\label{ssec:blueprint}
Constructing a web authentication scheme that fulfills one-out-of-two security and is thus only vulnerable to attackers that are able to compromise both devices is not an easy task.
As we will show in \cref{sec:related-work}, commonly used schemes do not have this property.

We propose \ac{2DA-Blueprint}, a blueprint for secure web authentication.
Following the blueprint simplifies constructing schemes that fulfill one-out-of-two security. 
We identify the following requirements for \ac{2DA-Blueprint}.

\begin{enumerate}[label=(2DA.\theenumi), leftmargin = *]
	\item \label{2DA_rule_ta} Authenticate each security-critical transaction individually.
	\item \label{2DA_rule_twodevice} Require verification of transaction details by the user on both devices.
	\item \label{2DA_rule_securemechanism} Use mechanisms for client and server authentication that protect against non-malware attacks.
	\item \label{2DA_rule_limitaccess} Use local authentication mechanisms to limit access to the devices.
\end{enumerate}

The first requirement addresses that the security guarantees can only be stated for transactions that are authenticated individually.
The second requirement is necessary to thwart attacks that involve one of the two devices being compromised.
Entering the transaction details on a device instead of reading them from the display is also possible.
Furthermore, the third requirement relates to the communication between the devices and the server.
It ensures that an attacker is not able to authenticate to the service without compromising a device.
For example, using a password for login, compromise of the primary device is not necessary to attack the scheme.
Password and phishing attacks are sufficient to bypass authentication of the primary device.
Thus, one-out-of-two security cannot be fulfilled by relying on password authentication for one of the devices.
Finally, the fourth requirement protects devices in case they are stolen.
Even though this is not strictly necessary to fulfill one-out-of-two security, we consider it important as it prevents physical attackers from compromising or abusing stolen devices.

Note that a secure transaction authentication scheme does not protect a user from authenticating an unintended transaction if the original transaction data was manipulated in the first place.
Thus, it is of utmost importance that the transaction data is either self-explanatory or the user has the ability to compare the transaction data with a known safe value.
For example, attacks on online banking have been described that are based on manipulating digital invoices~\cite{haupert2019short}.
If the original transaction details are only available on one device, then no scheme can offer protection when the device is compromised.

The goal of \ac{2DA-Blueprint} is to simplify creating web authentication schemes that protect security-critical transactions by relying on \emph{one-out-of-two security}.
In the following section, we show that building on existing web authentication mechanisms such as \ac{FIDO2}, we can use this new paradigm to design an authentication scheme that is secure, even in the presence of malware and real-time phishing attacks.
\newcommand{\compFormat}[1]{\textsf{#1}\xspace}
\newcommand{\User}{\compFormat{U}}
\newcommand{\Browser}{\compFormat{B}}
\newcommand{\Server}{\compFormat{S}}
\newcommand{\PushServer}{\compFormat{PS}}
\newcommand{\App}{\compFormat{A}}

\section{Two Display Authentication using \acs{FIDO2}}
\label{sec:newscheme}
We use our blueprint \ac{2DA-Blueprint} from the last section to design \ac{2DA-Scheme}, a new scheme for web authentication that fulfills one-out-of-two security, i.e., the scheme is secure as long as one of the two devices is not compromised.
We refer to \cref{sec:newscheme-proof} for a detailed security proof.
On a high level, a user \User aims to initiate a transaction at a server \Server using a computer \Browser running a browser.
The user also holds an additional device \App, such as a smartphone.

We assume that \ref{2DA_rule_limitaccess} is already satisfied by existing mechanisms that restrict the access to \Browser and \App.
Following \ref{2DA_rule_securemechanism}, we build on top of an existing authentication mechanism with appropriate security, and following \ref{2DA_rule_twodevice} we use two instances of this mechanism.
That is, one instance is executed between \Browser and \Server, and one instance is executed between \App and \Server.
Again, following \ref{2DA_rule_twodevice} we let the user verify the transaction details during both instances.
To be more precise, we organize both instances in a way that the user enters the transaction details in the browser on \Browser and require verification of the transaction data by the user on \App using a confirmation channel.
In the following, we introduce the underlying authentication scheme \ac{FIDO2}, and then describe the protocol for registration and authentication in its entirety.

\subsection{\acs{FIDO2}} We use \ac{FIDO2} as the underlying authentication mechanism. That is, our protocol consists of two instances of \ac{FIDO2}. 
\ac{FIDO2} is an interactive public-key challenge-response authentication protocol published by the \ac{FIDO} Alliance.
It is used on top of TLS. Both registration and authentication consist of a three-message flow between client and server.
The first message from client to server initiates the interaction.
The second message is sent from server to client and contains a randomly sampled challenge.
Finally, the client sends a signature of this challenge and some additional data to the server.

More precisely, the client does not do all the computation itself, but rather forwards messages to a so-called authenticator, which contains all key material and performs cryptographic operations.
\ac{FIDO2} differentiates platform authenticators that are integrated into devices and roaming authenticators that can be connected via transports such as USB.
For our further description, we focus on the \ac{FIDO} client, because most authenticators do not contain a display to show transaction data themselves that is crucial for verification by \User.

There are mainly two reasons for our choice of \ac{FIDO2}.
First, it is supported as an API by all modern browsers on the client side and libraries for various programming languages on the server side. This makes our protocol easy to implement and integrate.
Second, it offers the security we need for \ref{2DA_rule_securemechanism} as well as the user verification we need for \ref{2DA_rule_limitaccess}.
Intuitively, \Ac{FIDO2} is resilient to replay and phishing attacks because both a random challenge as well as the identifier of \Server, typically a domain name, is signed.
Next, we describe our protocol \ac{2DA-Scheme} in its entirety to show how the two instances of \ac{FIDO2} work together.

\begin{figure}[!t]
	\centering
	\includegraphics[width=\linewidth]{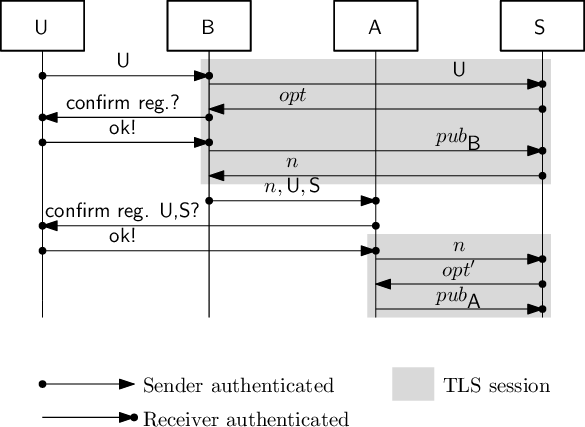}\hfill
	\caption{Sequence diagram for registration in our protocol \ac{2DA-Scheme}.
		Here, a user \User registers at a server \Server.
		The devices \Browser and \App are used to generate secret and public keys $s_\Browser,pub_\Browser$ and $s_\App,pub_\App$ during the process. The nonce $n$ is randomly generated by \Server.}
	\label{sequence_diagram:registration}
\end{figure}
\subsection{Registration}
First, the user registers \Browser using the standard \ac{FIDO2} registration ceremony.
During registration, \Server stores the public key $pub_\Browser$ from \Browser.
Then, \Server provides a nonce to \Browser to link the second device \App to the account.
The nonce is transferred from \Browser to \App using a QR code.
Again, we use the \ac{FIDO2} registration ceremony to store the public key $pub_\App$ from \App on \Server.

In more detail, the registration process is depicted in \cref{sequence_diagram:registration}.
The server responds to a user registration request with a set of options $opt$, containing a random challenge, the identifier of the server and the new user, as well as further parameters for the creation of credentials.
First, the authenticator of \Browser creates a new credential and sends the public key $pub_\Browser$ back to \Server.
During this process, the user is asked to confirm registration.
Next, the server \Server randomly generates a nonce $n$, links it to the username, and sends it to \Browser.
Then, $n$ is transferred from \Browser to \App (e.g., using a QR code).
Again, the user is asked to confirm the registration on the device \App.
Then, \App can initiate a similar \Ac{FIDO2} registration ceremony, allowing \Server to link both ceremonies using $n$.

As a result of the registration, \Server is aware of public keys $pub_\App, pub_\Browser$ linked to user \User.
The devices \App and \Browser know the secret keys $s_\App, s_\Browser$ for the public keys $pub_\App, pub_\Browser$, respectively.

\begin{figure}[!t]
	\centering
	\includegraphics[width=\linewidth]{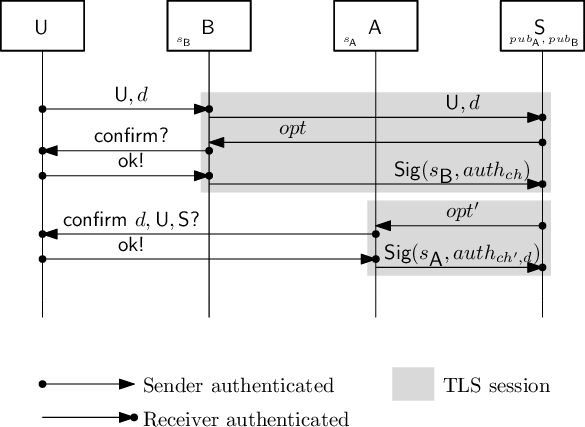}\hfill
	\caption{Sequence diagram for transactions in our protocol \ac{2DA-Scheme}.
		Here, a user \User authenticates a transaction with data $d$ with a server \Server by using devices \Browser and \App.
		The devices \Browser and \App have a secret key $s_\Browser$ and $s_\App$, and the server \Server knows the corresponding public keys $pub_\Browser$ and $pub_\App$.
		The variables $auth_{ch}$ and $auth_{ch',d}$ contain $ch$ and $ch',d$ respectively and are further explained in the text.}
	\label{sequence_diagram:transaction}
\end{figure}

\subsection{Transactions}
The message flow during a transaction is shown in \cref{sequence_diagram:transaction}.
Recall that a user \User wants to issue a transaction with data $d$ at a server \Server.
First, \User uses \Browser to transfer the intended transaction data $d$ in combination with the username to \Server.
\Server sends a set of options $opt$, containing a random challenge $ch$, and expecting a signature of $ch$ for the public key $pub_\Browser$ as a result.
Again, the user has to confirm before the challenge is signed by \Browser and the signature is sent to \Server.
To be more precise, the signature is computed over the challenge, an identifier of \Server, extension data, and the authenticator's signature counter, which is used for clone detection.
This follows the \Ac{FIDO2} standard.
In \cref{sequence_diagram:transaction} this is simply presented as $\mathsf{Sig}(s_\Browser,auth_{ch})$ while $auth_{ch}$ contains the aforementioned data.
After receiving the signed challenge from \Browser, the transaction is not yet fully authenticated.
The server \Server requires a signature from \App too.
After transmitting $d$ to \App, the same \Ac{FIDO2} protocol is performed between \App and \Server.
Note that for this protocol a fresh challenge $ch'$ and the public key $pub_\App$ is used.
\Server links both challenges $ch$ and $ch'$ to the transaction data $d$.
We use the \ac{txAuthSimple} of \ac{FIDO2} with data $d$ for \App. This extension ensures that \App receives the transaction data $d$ in combination with the challenge $ch'$ as part of options $opt'$, and presents it to the user.
After \User's confirmation, \App not only signs $ch'$, but also $d$.
This allows to bind the challenge $ch'$ to the transaction data $d$, which is important when the data is transmitted over an insecure channel.
As we propose to use TLS, transferring $d$ separately and not signing it would suffice in our attack model. However, using \ac{txAuthSimple} allows adapting the protocol to devices that are not connected to the Internet directly and thus require another device to relay the messages to them.
Again, there is additional data that is signed as in $auth_{ch}$, and we denote this as $auth_{ch',d}$ in our diagram.
\Server only accepts the transaction with data $d$ when both authentication ceremonies are successful and \App signed the correct transaction data $d$.

\subsection{Prototypical Implementation}
To verify the feasibility of our approach, we implemented a prototype of \ac{2DA-Scheme} consisting of a server component for \Server and an Android app running on \App.
On the server, we use a Go library for \ac{FIDO2} by Duo-Labs\footnote{\url{https://github.com/duo-labs/webauthn}}.
We had to add support for the \ac{txAuthSimple} extension.
More specifically, the server has to check that the authenticator signed the extension data containing the transaction details and that they have not been tampered with.
For our app, we use an Android library by Duo-Labs\footnote{\url{https://github.com/duo-labs/android-webauthn-authenticator}} to create credentials and sign supplied challenges.
Again, we added support for the \ac{txAuthSimple} extension, which mainly required signing the extension data.
On device \Browser, we use the \Ac{WebAuthn} browser API which is part of the \ac{FIDO2} standard and is supported by all modern browsers.

Following the principle~\ref{2DA_rule_limitaccess} we limit access to the devices by a local authentication mechanism such as a lock screen.
On \App we force the use of a lock screen by enabling the option \emph{setUnlockedDeviceRequired} of the Android Keystore\footnote{\raggedright\url{https://developer.android.com/reference/android/security/keystore/KeyGenParameterSpec.Builder}}.
This ensures that the private key can only be used if the device is unlocked.
Access to the private key can be further restricted by requiring to push a physical button or displaying transaction data on a protected screen~\cite{mayrhofer2021android}.
However, these mechanisms might affect usability negatively and are not necessary to fulfill one-out-of-two security.
We assume that device \Browser has a lock screen set up too.
Furthermore, we enable the \emph{requireUserVerification} parameter in the \ac{WebAuthn} API and require the \emph{User Verified} flag to be set in the signed authenticator data.
This ensures that roaming authenticators that are prone to theft cannot be used without authentication e.g., using a PIN.

\begin{figure}[t]
	\centering
	\fbox{\includegraphics[width=.45\linewidth]{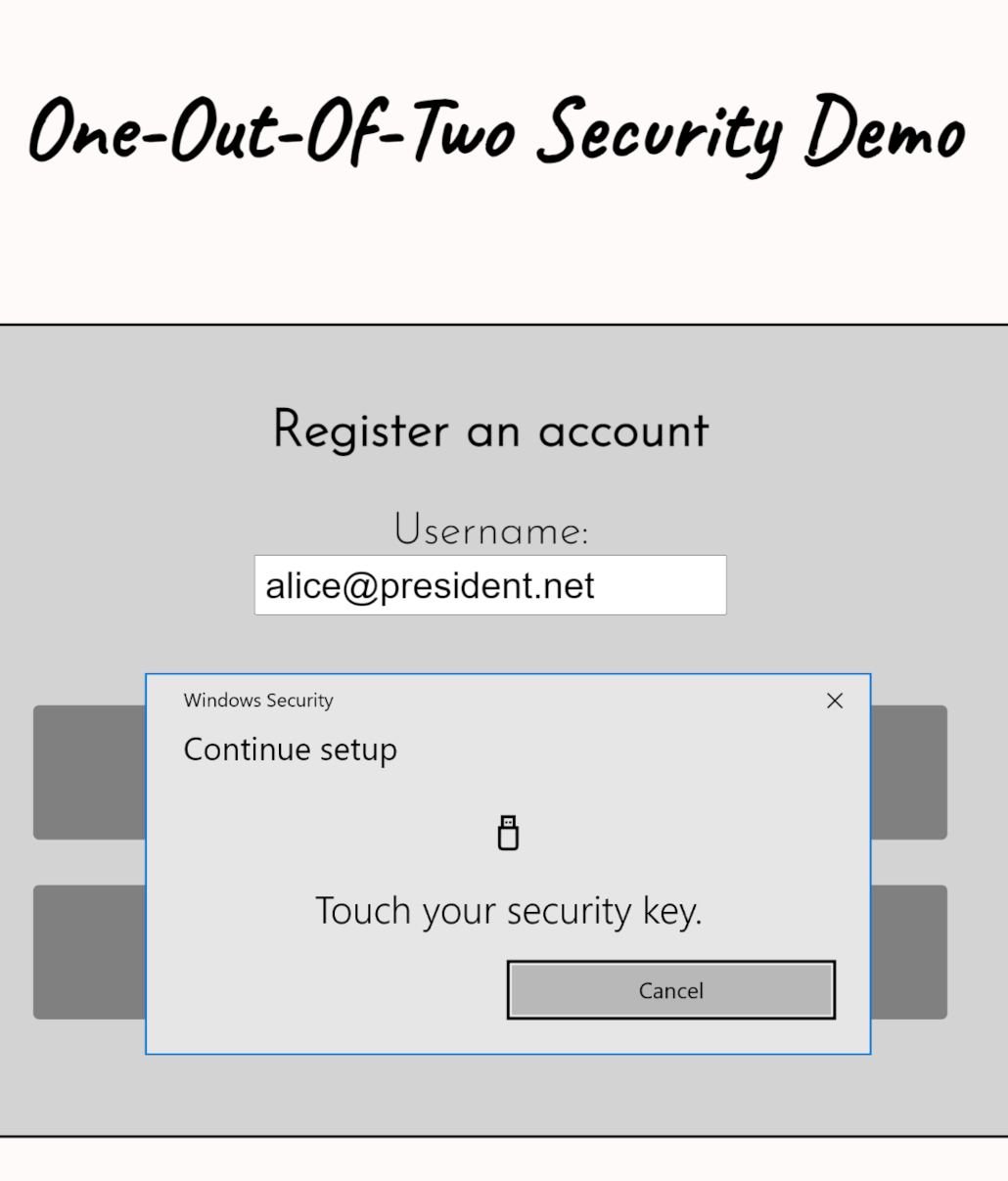}}\hfill
	\fbox{\includegraphics[width=.45\linewidth]{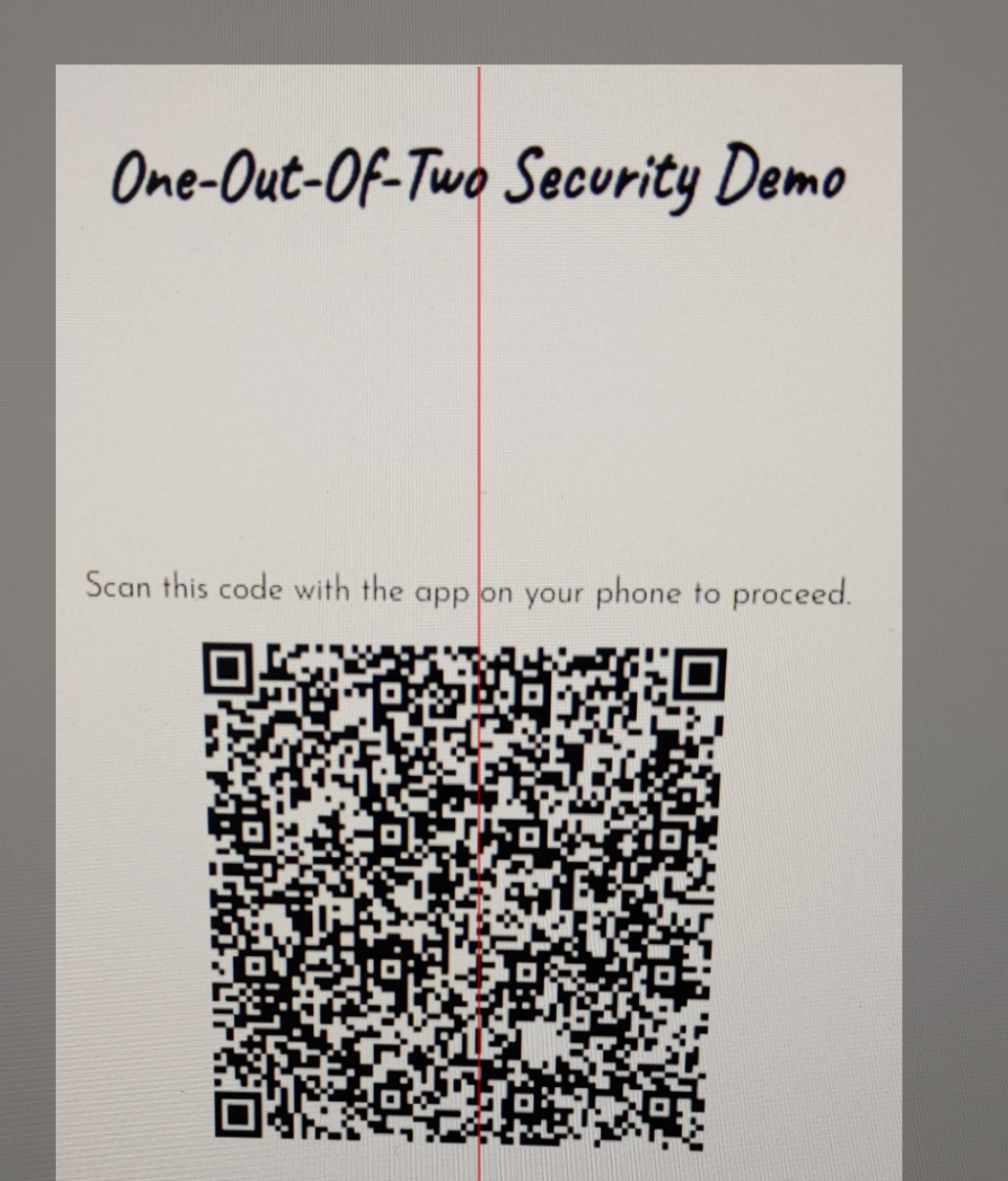}}\hfill
	\caption{Screenshots of a user's view in our app prototype during registration.
		\textit{Left:} Registration in the browser using \ac{FIDO2}.
		\textit{Right:} Registration in the app by scanning a QR code.}
	\label{fig:screenshots_reg}
\end{figure}

\begin{figure}[t]
	\centering
	\includegraphics[width=.48\linewidth]{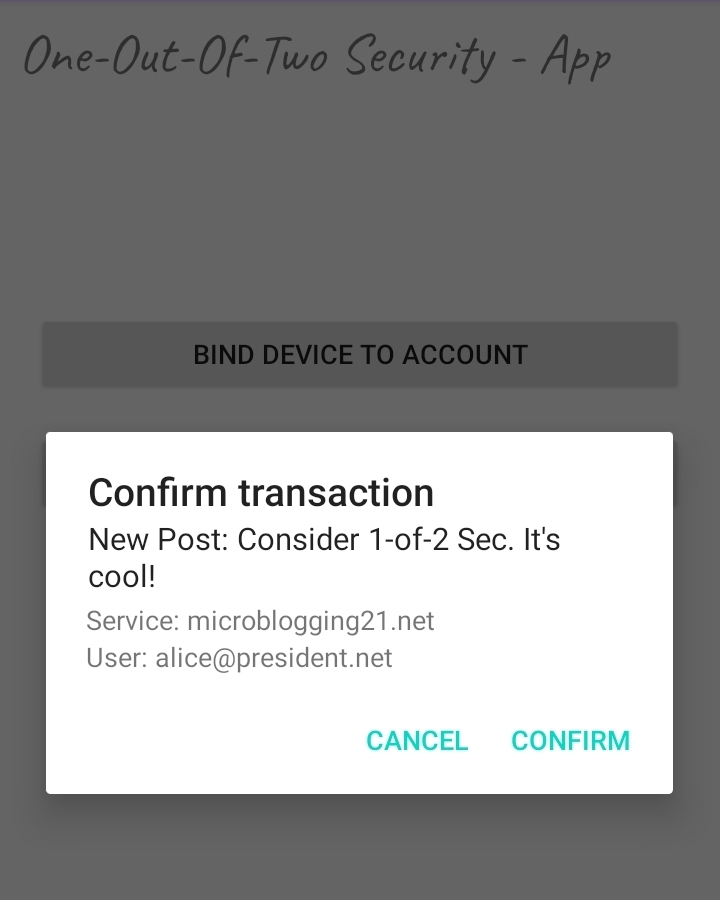}\hfill
	\caption{Screenshot of our app prototype during transaction confirmation in a fictitious micro-blogging scenario.}
	\label{fig:screenshots_auth}
\end{figure}

\cref{fig:screenshots_reg} shows the user interface of the registration process.
On \Browser the browser shows a pop-up asking the user to confirm registration with \ac{FIDO2}.
Afterwards, the user links the app on \App to the account by scanning a QR code from the browser.
In the background, the app establishes a connection to the server and initiates a \ac{FIDO2} registration ceremony as well.
\cref{fig:screenshots_auth} shows the transaction confirmation screen of our app.
In this example, the user initiated a new post on a fictitious micro-blogging service.
The app displays all relevant information including the username and identifier of the service.

Even though the \ac{CTAP} protocol is designed to relay data to an authenticator, we decided against using it for our App.
\ac{CTAP} requires the user to connect the smartphone to the computer during authentication~\cite{ctap}.
However, Bluetooth and NFC are not available on all platforms and USB requires a wired connection.
Furthermore, current browsers do not forward unsupported extensions and thus \ac{txAuthSimple} could not be used~\cite{klieme2020fidonuous}.
Thus, we suggest using push messages and TLS for the communication of \App and \Server.
In our implementation, we use the push service Firebase Cloud Messaging\footnote{\url{https://firebase.google.com}}.
However, we do not transmit any data in the push message, but request the app to establish a TLS connection to the server and retrieve data from there.
Integration of a push-based transport into \ac{CTAP} would be desirable, as it would allow \ac{2DA-Scheme} to rely on standard \ac{FIDO2} only.
Recently, attempts to integrate such a mechanism into \ac{CTAP}~\cite{ctapduo2020} have been made, however, they are not yet included in the standardized protocol.
We hope that our work strengthens these developments.
Furthermore, our prototypical app could be superseded by integrated support for \ac{FIDO2} in mobile operating systems.
This requires that the smartphone can be used as a roaming authenticator on another device and that it supports the \ac{txAuthSimple} extension such that transaction data can be verified.
Sadly, the extension \acs{txAuthSimple} has been removed in the second level of the \ac{WebAuthn} API~\cite{webauthn}.

Although we do not consider recovery in our implementation, promising methods for recovery of \ac{FIDO2} authenticators have been proposed that do not sacrifice security by incorporating a backup authenticator~\cite{frymannasynchronous}.
\section{Security Proof}
\label{sec:newscheme-proof}
We formally prove the security property \emph{one-out-of-two-security} (see \cref{one-out-of-two-security}) of our protocol \ac{2DA-Scheme} with the help of \emph{\ac{tamarin}}~\cite{tamarin-prover}.
Our lemma definitions, which from \ac{tamarin}'s point of view constitute the protocol's security properties, are inspired by~\cite{konoth2020securepay}.
We also analyze the effect of users not verifying transaction data on the security of \ac{2DA-Scheme}.
While our scheme does not fulfill \emph{one-out-of-two-security} in this scenario, we prove that it still protects against most types of attacks including phishing and malware on the additional device.

In the following, we give a brief introduction to \ac{tamarin} and then describe how we modeled our protocol as well as the security property.

\subsection{A Short Introduction to Tamarin}
By modeling a protocol in its custom modeling language, \ac{tamarin} performs unbounded verification of given security properties (called \emph{lemmas}) for the protocol in the symbolic model.
Within the custom modeling language, individual protocol steps are expressed as \emph{multi-set rewrite rules}.
These rules describe world state changes during execution flow of a protocol.
This world state is tracked by \ac{tamarin} in the form of a set of \emph{facts}.
A fact is an atom, optionally associated with data such as key material, nonces, identities or constants.
A rule can query facts as well as add new facts to the state.

Application of a rule can be subject to the existence of a fact within the state.
Such a fact is called an \emph{in-fact}.
Similarly, facts that a rule adds to the global state, as a result of its application, are called \emph{out-facts}.
Chaining of two rules \trl{first}, and \trl{second} is achieved by matching \trl{first}'s out-facts with the in-facts of \trl{second} (see \cref{fig:tamarin-example-chain}).
Additionally, facts may contain data, such as cryptographic key material, constants and nonces.

\begin{figure}
\begin{lstlisting}[language=tamarin,breaklines=true,frame=single,basicstyle=\small]
rule first:
	[ ]
  	--[ ]->
	[NowSecond()]
	
rule second:
	[NowSecond()]
  	--[ ]->
	[NowThird()]
\end{lstlisting}
\caption{Example of chaining two \ac{tamarin} rules.}
\label{fig:tamarin-example-chain}
\end{figure}

Within \ac{tamarin}'s input language, interaction with the attacker is modeled by special in- and out-facts, which we briefly explain in the following:

\begin{itemize}
	\item
		\lstinline[basicstyle=\small]"Fr(~var)" is an in-fact that instantiates a variable binding that represents a perfect randomly chosen value.
	These values can (among other) be used to model generation of nonces and cryptographic key material.
	\item
		\lstinline[basicstyle=\small]"In(var)" is an in-fact that reads a value from the network.
	\item
		\lstinline[basicstyle=\small]"Out(var)" is an out-fact that writes the contents of its argument to the network.
\end{itemize}

These facts can be employed arbitrarily by the protocol's rules, are not captured in the global state and as such do not need to have a corresponding counterpart embodied in the global state already.

Facts within the global state as well as facts that are written on the trace, including their respective data captures, are not known to the attacker.
Instead, the attacker's knowledge is specifically extended by writing data to the attacker-controlled network.
This mechanism can be used to leak any kind of data such as key material, identities, nonces as well as a protocol's message flow.

A protocol declaration in \ac{tamarin} usually contains at least one rule with an empty set of in-facts.
This rule can be instantiated on the state that only contains facts that are populated by \ac{tamarin} at the start of a program verification session.
\subsection{Our Tamarin Model for \ac{2DA-Scheme}}
\begin{figure*}[t]
	\centering
	\includegraphics[width=0.8\linewidth]{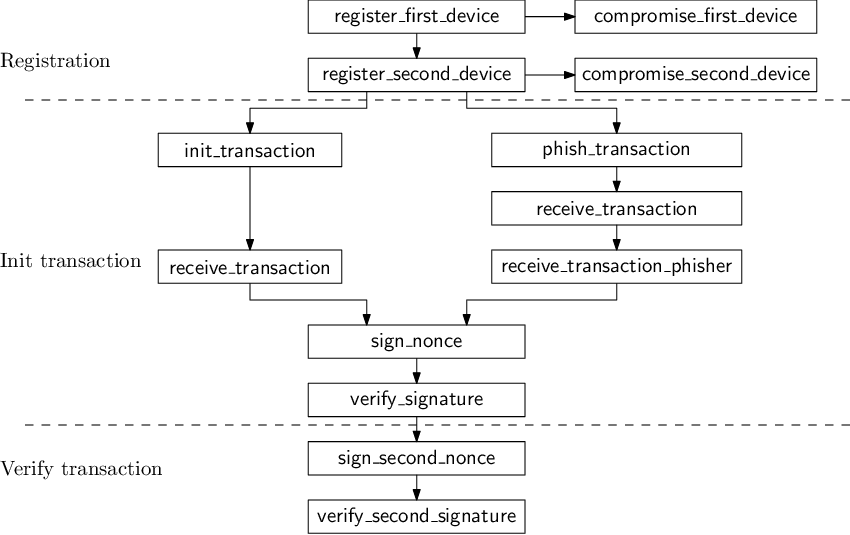}\hfill
	\caption{Structure of our Tamarin model. Pictured is the most straightforward order of instantiation of rules for normal and phishing transactions.}
	\label{fig:proof:tamarin-model}
\end{figure*}

Our model consists of 13 rules and two lemma definitions.
We refer to \cref{app:tamarin-proof} for a complete definition of each rule, including an enumeration of facts that are produced and consumed by it.
Each rule's purpose is briefly explained below:

\begin{itemize}

\item
	\trl{new\_server}:
	registers a new honest server as a global fact.
	Honest servers cannot be used in phishing attacks.

\item
	\trl{register\_first\_device}:
	Registers a new account at a server and connects a user's first device with it.
	The device is identified by a public key generated for the particular account.
	Account and device identities are then leaked to the attacker. 
	
\item
	\trl{register\_second\_device}:
	Finishes registration of a user's account by connecting a second device with it.
	This step also leaks the public key registered by the second device to the attacker.
\item
	\trl{init\_transaction}:
	A user commences a transaction at a server for which there already exists a personal account with two registered devices.
	After sending the transaction with its accompanying data to the network, it waits for a nonce from the server.
	
\item
	\trl{receive\_transaction}:
	An honest server receives a transaction from a previously registered user.
	It generates a nonce and sends this back to the user.

\item
	\trl{phish\_transaction}:
	A user initiates a transaction on a phishing server.
\item 	
	\trl{receive\_transaction\_phisher}:
	A phishing server receives a transaction of a previously phished user.

\item
	\trl{sign\_nonce}:
	A user signs a nonce on her first device for a transaction that has also been started by her.

\item
	\trl{verify\_signature}:
	A server verifies the signature of a previously issued nonce of a user's first device.
	It then generates a second nonce, which is sent to the user's second device. 

\item
	\trl{sign\_second\_nonce}:
	A user's second device receives a transaction request.
	After the user verified that the displayed transaction data corresponds to the transaction she previously initiated, her second device signs the second nonce.
	The signed nonce is sent back to the server.

\item
	\trl{verify\_second\_signature}:
	A user's second signature is verified on the server.
	If verification succeeds the transaction is completed, i.e., the server executes the user requested transaction internally.
	
\item
	\trl{compromise\_first\_device}:
	A user's primary device is compromised.
	We model this by leaking the device's private key to the adversary.

\item
	\trl{compromise\_second\_device}:
	A user's additional device is compromised.
	We model this by leaking the device's private key to the adversary.
	
\end{itemize}

An example trace containing a transaction accepted by a server can be obtained by instantiating the rules in the order given in \cref{fig:proof:tamarin-model}, excluding rules that model phishing or compromise of devices.
In our model, there are two rules that do not require custom facts to be present within the global state, namely \trl{new\_server} and \trl{register\_first\_device}.
Consequently, these two rules can be instantiated regardless of the current world state within \ac{tamarin}.
In this example, protocol instantiation starts with registration of an honest server (omitted for clarity in \cref{fig:proof:tamarin-model}), registration of a user's devices at this server, initialization of a transaction and verification of a transaction by the server.
These rules correspond to three distinct categories: \emph{registration}, \emph{transaction initialization} and \emph{transaction verification}.
In order to model phishing attacks, which is a transaction that is started on a malicious server, \emph{transaction initialization} contains two more rules, \trl{init\_transaction\_phisher} and \trl{receive\_transaction\_phisher}.

\begin{figure*}[t]
	\centering
	\begin{lstlisting}[language=tamarin,breaklines=true,frame=single,basicstyle=\small]
lemma only_user_initiated_transactions_accepted:
	
"All initiator transaction server #i. TransactionComplete(initiator, server, transaction) @i
==> 
((Ex #j. TransactionBegin(initiator, server, transaction) @j) | 
 (Ex #k #l. CompromiseDev1(initiator, server) @k & CompromiseDev2(initiator, server) @l))"
	\end{lstlisting}
	\caption{A \ac{2DA-Scheme} security property which states that only honest (i.e., user initiated) transactions are accepted by a server, captured as a \ac{tamarin} lemma. In conjunction with \emph{replay\_attack\_impossible}, this lemma constitutes \emph{one-out-of-two-security}.}
	\label{fig:honest-transactions}
\end{figure*}
\subsection{How Our Model Captures Reality}
\label{ssec:model_reality}
\ac{tamarin} builds on the Dolev-Yao\cite{dolev-yao} attack model.
In this model, the attacker obtains knowledge of every message sent over the network.
He is also able to modify message contents, suppress and inject messages as well as re-send known messages at any time in protocol execution.
However, the model assumes that the attacker is not able to break cryptographic schemes without knowing the key.

Although we expect typical implementations of \ac{2DA-Scheme} to be embedded within already bootstrapped \ac{TLS} sessions, our \ac{tamarin} model relies on unilateral authentic channels only.
Specifically, we require the communication channels from the server to the user's devices to be authentic.
We use the rules suggested in the \ac{tamarin} manual to model the authentic channel~\cite{tamarin-manual}.
As a result, the actual payload of a transaction is always assumed to be transmitted in the clear over an attacker-controlled network.
These simplifications allow us to present a more concise model in contrast to a model in which \ac{FIDO2} and \ac{TLS} have also been realized.
Furthermore, this highlights again that we assume a strong attack model.
In a real-world implementation of \ac{2DA-Scheme}, confidentiality is guaranteed by performing transactions within the context of a \ac{TLS} session.

\ac{2DA-Scheme} relies on \ac{FIDO2} and thus we also model a simplified version of the \ac{FIDO2} protocol.
We use the built-in \emph{signing} model of \ac{tamarin} that provides the necessary functions of a signature scheme.
To model drawing a fresh nonce as challenge, we use the built-in fact \lstinline"Fr(~nonce)".
We simplified the signed authenticator data to include the server identity and nonce only.
Thus in our model, the signature generation can be described as \lstinline"sign(<S, nonce>, privkey)".
In our implementation, the second device signs the transaction details by usage of the FIDO2 \ac{txAuthSimple} extension. 
Therefore, our model assumes that the authenticator data also contains the transaction data for the second device.
This is expressed by the term \lstinline"sign(<S, d, nonce>, privkey)".

During registration, the registered public keys are written to the attacker-controlled network.
We thereby model that authentication information stored on a server can be leaked by a data breach.
However, user devices must not be compromised during registration.

Our model considers phishing and malware attacks as described in \cref{sec:problems}.
We model malware by explicitly allowing the attacker to compromise a device.
This is depicted in the rules \trl{compromise\_first\_device} and \trl{compromise\_second\_device} by leaking the stored private key.
Leaking the private key is a worst-case scenario, as the private key is sufficient to impersonate a device completely.

To consider real-time phishing, we add the rules \trl{phish\_transaction} and \trl{receive\_transaction\_phisher}.
The rule \trl{phish\_transaction} allows the attacker to force the user to visit a potentially attacker-controlled website.
In an actual attack, this could be accomplished by sending a phishing email with a link to the user.
We assume that the user does not notice the phishing attempt and continues to follow the protocol by entering transaction data.
The rule \trl{receive\_transaction\_phisher} allows the attacker to answer the user's request with arbitrary transaction data and nonce.
The whole protocol flow of a phishing attack can be found in \cref{fig:proof:tamarin-model}.
After receiving transaction data from the user, the attacker initiates a manipulated transaction at the original server.
For example, this allows him to relay the nonce of a different transaction to the user.

\subsection{Proving One-out-of-two Security}
This model is verified by \ac{tamarin} against two lemmas, \emph{only-user-initiated-transactions-accepted} and \emph{replay-attacks-impossible}.
Together they form our protocol's notion of \emph{one-out-of-two-security}.
The lemmas are inspired by the \ac{tamarin} proof of SecurePay~\cite{konoth2020securepay}.
We give the definition of the first lemma in \cref{fig:honest-transactions}.
For brevity, we omit the second lemma and refer to \cref{app:tamarin-proof} for the complete definition.
Informally the first lemma states that every transaction that is accepted by a server has been initiated by an honest user and has not been tampered with.
Additionally, the second lemma is necessary to prevent replay attacks.
It ensures that each accepted transaction corresponds to exactly one transaction initiated by an honest user.
Furthermore, both lemmas contain an additional clause.
The additional clause ensures that the lemma is also satisfied if \emph{both} of the user's devices have been compromised.
In this case, the protocol cannot provide any security guarantees.

For simplicity, the lemmas do not impose any temporal order on the sequence of actions on the trace.
Thereby, the lemmas are more general but consider traces with specific temporal order as well.
For example, the point of time at which the start of a transaction is recorded should be before the corresponding completion of said transaction.

By using the \ac{tamarin} theorem prover, we verified that our protocol satisfies both lemmas \emph{only-user-initiated-transactions-accepted} and \emph{replay-attacks-impossible} and thus exhibits one-out-of-two security.
We sketch why the scheme intuitively fulfills one-out-of-two security and adheres to these lemmas.
Assume a user's first device is compromised and her second device is benign.
Then, the attacker is able to issue fraudulent transactions or manipulate benign transactions.
During protocol execution the user is asked to confirm the transaction data $\hat{d}$ on her second device although she did not initiate a transaction at all or initiated a transaction with data $d \neq \hat{d}$.
Clearly, the user will not confirm this transaction and thus the server will not execute the transaction.
On the other hand, assume her second device compromised and her first device benign.
In this case, the attacker cannot initiate a transaction himself.
If the user initiates a transaction, the attacker is not able to manipulate the transaction data because the server links the challenge sent to the user's second device to the transaction data entered on the corresponding first device beforehand.
Thus, the attacker can only complete a fraudulent transaction by compromising both devices.

\subsection{Comparing Transaction Data}
\label{ssec:compare}
Our model expects the user to verify transaction data properly.
User studies suggest that this is not always the case~\cite{haupert2019look}.
Similar to prior work, we extend our model to consider users that do not verify transaction data at all~\cite{basin2016modeling}.
In our model, it is sufficient to remove parameters from the fact \lstinline"UserWaitForConfirmation".
Thereby, the user only confirms a transaction after she initiated one but does not compare the transaction data.
Under this assumption, our protocol does not satisfy \emph{one-out-of-two-security}.
An attacker can break the scheme by compromising the first device to manipulate a transaction initiated by the user.
Then, the user confirms the manipulated transaction on the second device because she does not verify the transaction data.
However, by removing the rule \trl{compromise\_first\_device} we can prove that our protocol resists an attacker that compromises the additional device and carries out password and phishing attacks even if the user does not verify the transaction details at all.
Thus, \ac{2DA-Scheme} provides strong security guarantees and even provides \emph{one-out-of-two-security} if the user verifies transaction data.

\section{On When To Use FIDO2D}
\label{sec:usecases}

\ac{2DA-Scheme} can improve security in several scenarios.
Before we describe possible use cases, we present general guidelines and requirements.

\subsection{Guidelines and Requirements}

To adopt \ac{2DA-Scheme}, the first step is to identify security-critical actions that should be protected from malware attacks.
Our scheme can be used to achieve two separate security goals.
First, the \emph{confidentiality} of data stored on the server can be protected by requiring the user to confirm requests for said data.
However, once the user approves a request, the data is shown on the primary device and might be eavesdropped by malware.
Nonetheless, malware on a single device cannot get access to data protected by transactions without the user requesting or confirming it.
Furthermore, transactions can protect the \emph{integrity} of data stored on the server or activities triggered by the server.
For example, money transfers are often protected by transactions as they change the account balance stored on the server.
Actions have to be chosen carefully, as requiring \emph{transaction authentication} for many actions can easily cause authentication fatigue, drastically reducing security benefits.

For the security of \ac{2DA-Scheme}, it is important that transaction details contain all necessary information about the transaction, so that the user can verify them.
As \ac{2DA-Scheme} might be used for multiple accounts and services, the transaction details should always include the domain name of the service and the account name.
Furthermore, the length of the transaction details should also be considered when identifying suitable actions.

Services can still provide a login mechanism and offer to initiate a transaction from the authenticated session.
However, security-critical actions chosen by the provider must be verified by the user on the additional device.
Of course, not all users want to confirm messages individually.
Thus, the use of \ac{2DA-Scheme} should be configurable.
However, disabling \ac{2DA-Scheme} for an account must also require verification with \ac{2DA-Scheme}.
Otherwise, \ac{2DA-Scheme} can be bypassed trivially.

To be able to use \ac{2DA-Scheme}, users need a smartphone and a computer that supports \ac{FIDO2} either with an integrated platform authenticator or with a roaming authenticator.
Windows 10 provides a \ac{FIDO2} platform authenticator out-of-the-box\footnote{\url{https://fidoalliance.org/microsoft-achieves-fido2-certification-for-windows-hello/}}.
Hence, many users only need to install an app on their smartphone.
Our app can be used for multiple web applications that implement our protocol.
While platform authenticators are bound to a specific device, a roaming authenticator might also be used to access an account from multiple devices.

\subsection{Use Cases}
\ac{2DA-Scheme} is suited best for scenarios with security-conscious users as it unfolds it full potential when users verify transaction data properly.
We introduce four use cases where \ac{2DA-Scheme} can protect users against malware attacks.

\subsubsection{Online Banking}
Of course, \ac{2DA-Scheme} can be used in online banking scenarios.
For example, transactions can be used to protect the \emph{integrity} of money transfers and settings such as transfer limits.
For money transfers, the transaction details should include the recipient and the amount.
Changing limits, should include the new value of the limit.
These types of transactions can be verified on a smartphone properly.

Even though \emph{transaction authentication} is already used in online banking, applying \ac{2DA-Scheme} yields the following advantages.
First, it encourages the user to initiate and confirm transactions on separate devices.
And second, it replaces the use of passwords to access the online banking system.

\subsubsection{Electronic health record}
In electronic health records, \ac{2DA-Scheme} can be used to protect the \emph{confidentiality} of stored health information.
Sensitive information would only be accessible for a patient after confirmation on the smartphone.
The transaction data should indicate which data was requested.
Such requests can be verified on a smartphone properly.

\subsubsection{Microblogging}
In a microblogging service, the \emph{integrity} of posted messages can be protected using \ac{2DA-Scheme}.
In this scenario, publishing a message might have a huge impact depending on the author.
Messages of influential people such as the former president of the United States have been associated with stock market activity~\cite{gjerstad2021president} but might also influence international affairs.
As messages in a microblogging service are short (e.g., 280 characters on Twitter), they are suitable for verification on a second device.
The transaction details should include the name of the account as well as the full message content.

\subsubsection{Administration Panels}
Management panels such as Plesk\footnote{https://www.plesk.com/} allow users to carry out administrative tasks on remote servers.
For example, DNS settings for registered domains can be configured.
Manipulating the DNS settings of a domain allows to eavesdrop and manipulate all traffic destined to the domain~\cite{mandiant2019dnshijack}.
Thus, protecting \emph{integrity} of the DNS settings using \ac{2DA-Scheme} is beneficial.
The transaction details should include the domain name, the record type and the value.
Since this information is short, it is suitable for verification on a smartphone.

\section{Related Work}
\label{sec:related-work}

We consider three types of related work: formal frameworks to prove security properties of cryptographic protocols, security models and work that introduces web authentication schemes.
Finally, we compare the security of existing web authentication schemes with \ac{2DA-Scheme}.
\subsection{Proof Frameworks}
Defining and proving the security of a protocol within a formal model is an important step during the development of new security protocols.
To this end, various frameworks and models have been proposed in the past. 
The \ac{UC} Framework~\cite{uc2001} for example is widely used by the cryptographic community and is considered to be the most popular and most expressive framework.
It is best suited for modeling basic cryptographic primitives and simple protocols; however, modeling protocols is a manual task and there is no support for automated proof checking.
Lots of technical artifacts (like its reliance on Turing Machines for modeling protocol participants) therefore make modeling complex protocols error-prone and lead to complex security proofs which even humans sometimes cannot verify completely.

In contrast, the \ac{tamarin} modeling framework allows verification of models in an automated fashion~\cite{tamarin-prover}.
\ac{tamarin}-models consist of \emph{multi-set rewrite rules} as well as lemmas, defining protocols and security properties respectively.
These constructs are expressed in a domain specific language.
Protocols and their security properties are verified in the symbolic model.
A proof of correctness can then be obtained by executing the \ac{tamarin} theorem prover.
A curated list of \ac{tamarin}-proofs for various protocols can be found on the \ac{tamarin} website\footnote{https://tamarin-prover.github.io/}.
For instance, \ac{tamarin} has been used to verify all claimed security properties of a draft specification of TLS 1.3~\cite{tls-13-tamarin}.

\subsection{Security Models}
To remedy the issue of a single point of failure within a trusted device, Achenbach et al.~\cite{achenbach2019your} put forth the notion of \enquote{one-out-of-two security}.
They provide a model for the security of electronic payment protocols based on the \ac{UC} framework.
Furthermore, they introduce a protocol for cash withdrawal at an ATM with an additional device such as a smartphone, and prove that it fulfills one-out-of-two security.
Thus, the protocol is secure as long as one device is not compromised.
While their protocol and model are specifically tailored to electronic payments and cannot be directly applied to web authentication, our work is inspired by theirs.

Jacomme et al. introduce a formal model for multi-factor authentication in the applied pi calculus~\cite{jacomme2021extensive}.
They analyze the \ac{U2F} protocol, a predecessor of \ac{FIDO2} and Google 2-step authentication.
Their threat model includes phishing and malware attacks, as well as fingerprint spoofing to bypass \ac{RBA}.
Using ProVerif they identify several weaknesses of the analyzed protocols.
Google 2-step is susceptible to phishing attacks that include fingerprint spoofing and \ac{U2F} to malware controlling an attached authenticator.
They propose that these protocols should incorporate verification of transaction data, which is exactly what \ac{2DA-Scheme} provides.

Bonneau et al.~\cite{bonneau2012quest} compare schemes proposed to replace passwords for web authentication regarding usability, deployability, and security.
They consider a wide range of schemes such as password managers, federated schemes, \acp{OTP}, and hardware tokens.
The security of a scheme is assessed based on the resilience to different kinds of attacks.
However, phishing and malware attacks are not considered to their full extent.
For example, malware is only assumed to steal credentials passively but not to manipulate the browser.
Furthermore, real-time phishing attacks are excluded explicitly.

\subsection{Web Authentication}
\begin{table*}[!t]
	\renewcommand{\arraystretch}{1.2}
	\caption{Security of transaction authentication schemes.\protect\\
		\checkmark indicates that a scheme is secure in a given scenario. $\times$ indicates that this is not the case.
		\protect\\\Browser is the device running the browser. \App is an additional device.
		\protect\\Compare means that the user verifies transaction details. No Compare indicates that this is not the case.
		\protect\\Devices listed in the security assumption of a scheme must not be compromised.
	}
	\label{tab:sec-schemes}
	\centering
	\begin{tabular}{l | c c | c c | c}
		& \multicolumn{2}{c|}{No Compare} &  \multicolumn{2}{c|}{Compare} &      \\
		& \Browser compromised & \App compromised & \Browser compromised & \App compromised & Security assumption \\ \hline
		CAP reader\cite{drimer2009optimised}     & $\times$ & $\times$ & $\checkmark$         & $\times$ & \App \\
		PhotoTAN\cite{achenbach2019your}         & $\times$ & $\times$ & $\checkmark$         & $\times$ & \App \\
		MP-Auth\cite{mannan2011leveraging}       & $\times$ & $\times$ & $\checkmark$         & $\times$ & \App \\
		Chow et al.\cite{chow2016authentication} & $\times$ & $\times$ & $\checkmark$         & $\times$ & \App \\
		Secure Pay\cite{konoth2020securepay}     & $\times$ & $\times$ & $\checkmark$         & $\times$ & TEE in \App \\
		\ac{2DA-Scheme}                      & $\times$ & $\checkmark$  & $\checkmark$         & $\checkmark$ & \App or \Browser \\
	\end{tabular}
\end{table*}

\ac{FIDO2} is a standard for web authentication published by the \ac{FIDO} Alliance.
It consists of a widely implemented browser API called \ac{WebAuthn}~\cite{webauthn}.
This API simplifies integration of strong authentication mechanisms in web applications.
The underlying challenge-response protocol relies on public-key cryptography.
Even though the protocol is resistant to real-time phishing, it does not prevent malware attacks~\cite{jacomme2021extensive}.

This also applies to Google's Advanced Protection program~\cite{google2021advancedprotection}.
It restricts login attempts to \ac{FIDO} tokens; however, they can still be abused by malware.
Other proprietary solutions such as Duo Push and Akamai MFA confirm login attempts using push messages~\cite{akamain2021krypton, duo2021push}.
However, they do not authenticate transactions individually and are thus susceptible to malware using session hijacking or transaction manipulation (see \cref{sec:problems}).

The \Ac{CAP} protocol is a transaction authentication scheme used in online banking~\cite{drimer2009optimised}.
A dedicated card reader is used to allow the user to verify transaction details.
Even though the \ac{CAP} protocol is resilient to malware on the primary device, it does not exhibit one-out-of-two security.
In the unlikely event that the card reader is compromised, the scheme can be attacked by using phishing to initiate a malicious transaction (see \cref{sec:problems}).
Nonetheless, it provides a high level of security as compromise of the card reader is less likely than of a smartphone.
However, the protocol is of proprietary nature and requires the user to carry an additional device, which has been shown to hinder adoption significantly~\cite{krol2015they}.

Recent attempts to improve authentication schemes focus on banning weak schemes but do not reason about security properties and requirements systematically, e.g., the European Commission issued the \ac{PSD2}, a regulation requiring \emph{strong customer authentication} for online banking~\cite{PSD22017}. \ac{PSD2} suggests that users have to be made aware of the transaction details, but does not prevent malware attacks, as it allows operating both factors on one device~\cite{haupert2018app}.
NIST introduced security levels for authentication called \ac{AAL}~\cite{grassi2017nist}, but does not systematically consider what needs to be authenticated. 
Thus, these regulations do not solve the problems of two-factor authentication schemes. 

Mannan and van Oorschot~\cite{mannan2011leveraging} introduce Mobile Password Authentication (MP-Auth).
The scheme is based on a trusted smartphone that stores cryptographic keys.
By incorporating public-key cryptography, the protocol achieves phishing-resistance.
The authors use \emph{transaction authentication} to thwart malware attacks, yet MP-Auth is only secure if the smartphone is not compromised.
Because smartphones are multi-purpose devices that are connected to the Internet, they are susceptible to malware as well.
Similarly, Chow et al. introduce a scheme that relies on a trusted smartphone~\cite{chow2016authentication}.
The smartphone is used as a \ac{OTP} generator that displays the transaction data and requires confirmation by the user.

Konoth et al. propose a solution to secure two-factor authentication when both factors are operated on a single smartphone~\cite{konoth2020securepay}.
Their scheme SecurePay uses \acp{OTP} and a trusted user interface to verify transaction data.
The authors formally model and verify the security of their scheme using \ac{tamarin}~\cite{tamarin-prover}.
By relying on a \ac{TEE}, Secure Pay is resilient to malware attacks.
However, this is only true as long as attackers do not exploit vulnerabilities in the \ac{TEE} implementation to escalate their privileges to the secure world, which has been successfully achieved in the past~\cite{cerdeira2020sok}.
Furthermore, \ac{TEE} implementations have been shown to be susceptible to fault injection and side-channel attacks~\cite{ryan2019hardware,qui2020voltjockey}.

Similarly to our work, they identify the issue that the current way two-factor authentication is defined and used is insecure and give guidelines on how to improve upon that.
However, they mainly focus on the fact that strict isolation of the two factors is needed which, even though it works in their case, in general is not sufficient to guard against malware attacks.
In contrast, our work gives a universally applicable guideline for how to design secure authentication schemes, which are resilient against malware.
In particular, our notion of one-out-of-two security does not require a special type of device and is agnostic to how isolation is achieved.

\subsection{Security of Transaction Authentication Schemes}
We summarize the security of web authentication schemes in~\cref{tab:sec-schemes}.
We only consider schemes that support \emph{transaction authentication}, because other schemes are susceptible to malware attacks in any case (see \cref{sec:paradigm}).
Because users might not verify transaction data properly, we differentiate two scenarios: \emph{compare} and \emph{no compare} (see \cref{ssec:compare}).
Furthermore, we assume the attacker to either compromise \Browser or \App.
In addition, the attacker may carry out password and phishing attacks (see \cref{sec:problems}).
A scheme is considered secure, if the attacker is not able to execute a malicious transaction.
As expected, none of the considered schemes provides security guarantees if both devices are compromised.

First, we analyze the security of the schemes when a user confirms transactions on an additional device without comparing transaction data.
In this case, all considered schemes are vulnerable to malware on the primary device \Browser because the user does not detect transaction manipulation.
In contrast to other schemes, \ac{2DA-Scheme} is secure if \App is compromised (see \cref{sec:newscheme-proof}), because transaction initiation is protected by \ac{FIDO2}, which is resilient to password and phishing attacks.

If the user compares transaction data, all schemes protect against malware on the primary device \Browser.
Again, \ac{2DA-Scheme} is the only scheme that is secure if \App is compromised.
Control of device \App allows an attacker to tamper with transaction verification~\cite{haupert2018app}.
However, \ac{2DA-Scheme} protects transaction initiation with \ac{FIDO2} against phishing and password attacks.
The other schemes rely on password authentication (potentially including \ac{RBA}).
As described in \cref{sec:problems}, in this case the attacker can mount a transaction initiation attack.

Thus, \ac{2DA-Scheme} is the only scheme that fulfills one-out-of-two security.
For users that verify transaction data, it provides resistance to password, phishing, and malware attacks as long as one device is not compromised.
Even if the user does not verify transaction data at all, \ac{2DA-Scheme} is resilient to malware on device \App.
Thus, it provides a higher level of security than existing schemes regardless of whether transaction data is verified or not.
\section{Conclusion and Future Work}
In this work, we showed how to protect security-critical web transactions against attacks with malware. 
Current web authentication schemes do not offer this protection.

We identified requirements for such authentication schemes, namely one-out-of-two security and transaction authentication.
Web authentication schemes that fulfill these requirements protect security-critical transactions against malware attacks as long as one device is not compromised.

We introduced \ac{2DA-Blueprint}, a generic blueprint for designing web authentication schemes that fulfill one-out-of-two security.
Based on this blueprint, we designed and implemented a new web authentication scheme called \ac{2DA-Scheme}, which is applicable to a wide range of use-cases.
We proved the security of \ac{2DA-Scheme} using \ac{tamarin}.

By relying on protocols and APIs of the \ac{FIDO2} standard, \ac{2DA-Scheme} can be integrated into web applications easily.
We demonstrate this by creating a prototypical implementation.

Lastly, we urge the FIDO Alliance to fully embrace one-out-of-two security and transaction authentication as a design paradigm for secure web authentication in future versions of their standard.

\newpage
\bibliographystyle{IEEEtran}
\bibliography{IEEEabrv,SecureTransactions.bib}

\onecolumn
\appendix[Tamarin Model for \ac{2DA-Scheme}]
\label{app:tamarin-proof}
Our \ac{tamarin} model including the security lemmas can be found below.
For brevity, we omitted sanity lemmas, as well as the variations of the model for users that do no compare transaction data.
However, the full \ac{tamarin} model is available online\footnote{\url{https://tinyurl.com/2022-fido2d-tamarin}}.
\begin{lstlisting}[language=tamarin,breaklines=true,frame=single, numbers=left,basicstyle=\small]
theory fido2d
begin
builtins: signing

rule new_server:
	[	]
	--[ HonestServer($S) ]->
	[ !Honest($S) ]

rule register_first_device:
	[ Fr(~privkey) 	]
	--[ ]->
	[
		!Ltk_Dev1($I, $S, ~privkey),
		!Pk_Dev1($I, $S, pk(~privkey)),
		RegisteredPartially($I, $S),
		Out(<$I, pk(~privkey)>)
	]

rule register_second_device:
	[
		Fr(~privkey),
		RegisteredPartially(I, S)
	]
	--[ AccountRegistered(I, S) ]->
	[
		!Ltk_Dev2(I, S, ~privkey),
		!Pk_Dev2(I, S, pk(~privkey)),
		!Registered(I, S),
		Out(<I, pk(~privkey)>)
	]
	

rule init_transaction:
	[
		!Registered(I, S),
		!Honest(S)
	]	
	--[ TransactionBegin(I, S, $d) ]->
	[
		UserWaitForConfirmation(I, S, $d),
		Dev1WaitForNonce(I, S, $d),
		Out(<I, S, $d>)
	]

rule receive_transaction:
	[
		!Registered(I, S),
		!Honest(S),
		In(<I, S, d>),
		Fr(~nonce)
	]
	--[ TransactionReceived(I, S, d) ]->
	[
		ServerWaitForSignature(I, S, d, ~nonce),
		Out_A(S, I, <I, d, ~nonce>) 
	]

rule phish_transaction:
	[
		!Registered(I, S),
		In(P)
	]	
	--[ TransactionBegin(I, S, $d), Phisher(P) ]->
	[
		UserWaitForConfirmation(I, S, $d),
		Dev1WaitForNonce(I, P, $d),
		Phished(I, S, P)
	]

rule receive_transaction_phisher:
	[
		In(<d, nonce>),
		Phished(I, S, P)
	]
	--[ ]->
	[
		Out_A(P, I, <I, d, nonce>) 
	]

rule sign_nonce:
	[
		Dev1WaitForNonce(I, S, d),
		In_A(S, I, <I, d, nonce>),
		!Ltk_Dev1(I, S, privkey)		
	]
	--[ NonceSigned(I, S, nonce) ]->
	[
		Out(sign(<S, nonce>, privkey))
	]

rule verify_signature:
	[
		In(signature),
		ServerWaitForSignature(I, S, d, nonce),
		!Pk_Dev1(I, S, pubkey),
		Fr(~nonce2)
	]
	--[ Eq(verify(signature, <S, nonce>, pubkey), true), SignatureVerifiedDev1(I, S, d) ]->
	[
		ServerWaitForSecondSignature(I, S, d, ~nonce2),
		Out_A(S, I, <S, I, d, ~nonce2>) 
	]

rule sign_second_nonce:
	[
		UserWaitForConfirmation(I, S, d),
		In_A(S, I, <S, I, d, nonce>),
		!Ltk_Dev2(I, S, privkey)		
	]
	--[ DisplayData(I, S, d), NonceSigned(I, S, nonce) ]->
	[
		Out(sign(<S, d, nonce>, privkey))
	]

rule verify_second_signature:
	[
		In(signature),
		ServerWaitForSecondSignature(I, S, d, nonce),
		!Pk_Dev2(I, S, pubkey)
	]
	--[ Eq(verify(signature, <S, d, nonce>, pubkey), true), TransactionComplete(I, S, d)]->
	[
	]

rule compromise_first_device:
	[
		!Ltk_Dev1(I, S, privkey)
	]
	--[ CompromiseDev1(I, S) ]->
	[
		Out(privkey)
	]

rule compromise_second_device:
	[
		!Ltk_Dev2(I, S, privkey)
	]
	--[ CompromiseDev2(I, S) ]->
	[
		Out(privkey)
	]

// Authentic Channel Rules from Tamarin Manual
rule ChanOut_A:
	[
		Out_A(A, B, x)
	]
	--[ ChanOut_A(A, B, x) ]->
	[
		!Auth(A, x),
		Out(<A, B, x>)
	]

rule ChanIn_A:
	[
		!Auth(A, x),
		In(B)
	]
	--[ ChanIn_A(A, B, x) ]->
	[
		In_A(A, B, x)
	]

restriction Equality:
	"All x y #i. Eq(x,y) @i ==> x = y"

restriction HonestServersDontPhish:
	"All server #i #j. HonestServer(server) @i & Phisher(server) @j ==> F"

// This restriction is not required for security but simplifies proofs and counterexamples
restriction UniqueAccounts:
	"All acc server #i #j. AccountRegistered(acc, server) @i & AccountRegistered(acc, server) @j ==> #i = #j"

lemma only_user_initiated_transactions_accepted:
	"All initiator transaction server #i. TransactionComplete(initiator, server, transaction) @i ==> ((Ex #j.
		TransactionBegin(initiator, server, transaction) @j) | (Ex #k #l. CompromiseDev1(initiator, server) @k & CompromiseDev2(initiator, server) @l))"

lemma replay_attack_impossible:
	"All initiator1 transaction1 initiator2 transaction2 server #i #j. TransactionComplete(initiator1, server, transaction1) @i & TransactionComplete(initiator2, server, transaction2) @j & not #i = #j ==> ((Ex #k #l. TransactionBegin(initiator1, server, transaction1) @k & TransactionBegin(initiator2, server, transaction2) @l & not #k = #l) | (Ex #m #n. CompromiseDev1(initiator1, server) @m & CompromiseDev2(initiator1, server) @n) | (Ex #m #n. CompromiseDev1(initiator2, server) @m & CompromiseDev2(initiator2, server) @n) )"

end
\end{lstlisting}
\twocolumn

\end{document}